\documentclass[%
aps,%
preprint,%
showpacs,
nofootinbib,
groupedaddress]{revtex4}
\usepackage{epsfig}
\usepackage{amsmath}
\newcommand{\wt}{\widetilde}

\newcommand{\eqn}[1]{&\hspace{-0.7em}#1\hspace{-0.7em}&}

\newcommand{\abs}[1]{\mbox{$\left| #1 \right|$}}

\newcommand{\keV}{\mbox{keV}}
\newcommand{\GeV}{\mbox{GeV}}

\def\be{\begin{equation}}
\def\ee{\end{equation}}

\begin{document}

\title{The $\nu$MSM, Dark Matter and Baryon Asymmetry of the Universe}

\author{Takehiko Asaka}
\email{Takehiko.Asaka@epfl.ch}
\affiliation{
Institut de Th\'eorie des Ph\'enom\`enes Physiques,
Ecole Polytechnique F\'ed\'erale de Lausanne,
CH-1015 Lausanne, Switzerland}
\author{Mikhail Shaposhnikov}
\email{Mikhail.Shaposhnikov@epfl.ch}
\affiliation{
Institut de Th\'eorie des Ph\'enom\`enes Physiques,
Ecole Polytechnique F\'ed\'erale de Lausanne,
CH-1015 Lausanne, Switzerland}

%\date{\today}
\date{May 2, 2005}

\begin{abstract}

We show that the extension of the standard model by three 
right-handed neutrinos with masses smaller than the electroweak scale
(the $\nu$MSM) can explain simultaneously dark matter and baryon
asymmetry of the universe and be consistent with the experiments on
neutrino oscillations. Several constraints on the parameters of the
$\nu$MSM are derived.

\end{abstract}


\maketitle

%%%%%%%%%%%%%%%%%%%%%%%%%%%%%%%%%%%%%%%%%%%%%%%%%%%%%%%%%%%%%%%%%%%%%%%%
% Introduction
%%%%%%%%%%%%%%%%%%%%%%%%%%%%%%%%%%%%%%%%%%%%%%%%%%%%%%%%%%%%%%%%%%%%%%%%
{\em Introduction.}--- 
The canonical Minimal Standard Model (MSM) \cite{MSM}, despite being
extremely successful in particle physics, cannot accommodate
experimental data on neutrino oscillations ~\cite{nuexp} simply
because neutrinos are exactly massless in the MSM and thus do not
oscillate. In addition, the MSM does not contain any particle physics
candidate for cold dark matter and cannot explain the baryon
asymmetry of the universe.

The simplest renormalisable extension of the standard model,
consistent with neutrino experiments, contains  ${\cal N}$
right-handed SU(2)$\times$U(1) singlet neutrinos $N_I$
($I=1,\dots,{\cal N}$) with the most general gauge-invariant
interactions described by the Lagrangian:
\begin{eqnarray}
  \delta {\cal L}
  = \bar N_I i \partial_\mu \gamma^\mu N_I
  - F_{\alpha I} \,  \bar L_\alpha N_I \Phi
  - \frac{M_I}{2} \; \bar {N_I^c} N_I + h.c. \,,
  \label{lagr}
\end{eqnarray}
where $\Phi$ and $L_\alpha$ ($\alpha=e,\mu,\tau$) are respectively
the Higgs and lepton doublets, and both Dirac ($M_D = F \langle \Phi
\rangle$) and Majorana ($M_I$) masses for neutrinos are introduced.
We have taken a basis in which the mass matrices of charged leptons
and right-handed neutrinos are real and diagonal, and $F$ is a matrix
with elements $F_{\alpha I}$.

In addition to quite a large number of dimensionless Yukawa
couplings, this model contains ${\cal N}$ dimensionful parameters -
the Majorana masses of right-handed fermions. The neutrino
oscillation experiments cannot fix these new scales, as the masses
and mixing angles  of active neutrinos contain only specific
combinations of $M_D$ and $M_I$, coming from the diagonalization of
the complete mass matrix.

In this paper we propose to choose these unknown mass parameters to
be of the order of the electroweak scale or below. In other words, we
will assume that the model defined in (\ref{lagr}) is a true low
energy theory up to the Planck (or, say, grand-unified) scale.
Moreover, we fix ${\cal N}$ to be 3, keeping the number of
right-handed neutrinos to be equal to the number of fermionic
generations. The specific model in this parameter range will be
called below the ``$\nu$MSM", underlying its minimal character and
the fact that no new energy scale is introduced.

The aim of the present work is to demonstrate that the $\nu$MSM with
a particular choice of parameters, consistent with the data on
neutrino masses and mixing, can explain simultaneously the dark
matter and baryon asymmetry of the universe.

\newpage

{\em General properties of the $\nu$MSM.}--- 
The $\nu$MSM contains 18 new parameters in comparison with the MSM.
Three of them are the Majorana masses, while another 15 are hidden in
the Yukawa matrix $F_{\alpha I}$ and can be chosen as 3 diagonal
Yukawa couplings,  6 mixing angels and 6 CP-violating phases in the
following way:
\begin{eqnarray}
  F = \wt K_L \, f_{d} \, \wt K_R{}^\dagger \,,
\end{eqnarray}
where
\begin{eqnarray}
  f_d = \mbox{diag}(f_1, f_2, f_3) \,,~~~
  \wt K_L = K_L P_\alpha \,,~~~ \wt K_R{}^\dagger = K_R{}^\dagger P_\beta \,.
\end{eqnarray}
The diagonal matrices for Majorana phases are 
\begin{eqnarray}
  P_\alpha = \mbox{diag}( e^{ i \alpha_1 },  e^{ i \alpha_2 }, 1 ) \,,~~~
  P_\beta = \mbox{diag}( e^{ i \beta_1}, e^{i \beta_2}, 1 ) \,,
\end{eqnarray}
and the KM-like mixing matrix $K_L$ is given by
\begin{eqnarray}
  K_L ~\eqn{=}~
  \left(
   \begin{array}{c c c}
    1 & 0 & 0 \\
    0 &   c_{L23} & s_{L23} \\
    0 & - s_{L23} & c_{L23}
   \end{array}
  \right)
  \left(
   \begin{array}{c c c}
    c_{L13} & 0 & s_{L13} e^{-i \delta_L} \\
    0 & 1 & 0 \\
    -s_{L13} e^{i \delta_L} & 0 & c_{L13}
   \end{array}
  \right)
  \left(
   \begin{array}{c c c}
      c_{L12} & s_{L12} & 0 \\
    - s_{L12} & c_{L12} & 0 \\
    0 & 0 & 1
   \end{array}
  \right) \,,
\end{eqnarray}
where $c_{Lij} = \cos (\theta_{Lij})$ and $s_{Lij} = \sin
(\theta_{Lij})$.  The expression for $K_R$ is written in full analogy
with the replacement of $L$ to $R$. We fix the indices in such a way
that ``$1$" corresponds to the lightest right-handed neutrino and
``$3$" to the heaviest. The new light neutral fermions can naturally
be called the ``dark'' ($N_D$), for the dark matter
candidate,``clear'' ($N_C$) and  ``bright'' ($N_B$) sterile
neutrinos.

As the Majorana masses are assumed to be of the order of the
electroweak scale or below, the model can only be consistent with the
neutrino experiments if the Yukawa couplings are very small, $f_i^2
\sim O(m_\nu M_I/v^2)$, where $m_\nu$ are the masses of active
neutrinos and $v = 174$~GeV is the VEV of the Higgs field. At the
same time, in the interesting parameter range discussed below, the
Majorana masses are much larger than the Dirac masses, so that the
see-saw formula~\cite{Seesaw} for the masses of the active neutrinos
is applicable. The active neutrino mixing matrix \cite{numix} is
coming then from the diagonalization of the see-saw mass.

\newpage

{\em The $\nu$MSM and dark matter.}--- 
For small $f_i$ the lifetime of a right-handed neutrino may exceed
the age of the universe; in this case the $\nu$MSM provides a
particle physics candidate for the warm dark matter
\cite{SterileNeutrinoWDM}-\cite{Abazajian:2001nj}. The allowed mass
range of the corresponding sterile neutrino is severely restricted as
\begin{eqnarray}
  \label{eq:DMMass}
  2~ \keV \lesssim M_I \lesssim 5~ \keV \,,
\end{eqnarray}
where the lower bound comes from the cosmic microwave background and
the matter power spectrum inferred from Lyman-$\alpha$ forest
data~\cite{Viel:2005qj}, and the upper bound is given by the
radiative decays of sterile neutrinos in dark matter halos limited by
X-ray observations~\cite{Abazajian:2001vt}.

In the recent paper~\cite{Asaka:2005an} we have shown that the
minimal number of sterile neutrinos, which can explain the dark
matter in the universe, is ${\cal N}=3$.  In this case only one
sterile neutrino can be the dark matter. We identify it with the
lightest one $N_1$, i.e. $N_1$ is the ``dark'' sterile neutrino.  In
addition,  the observed mass density of the dark matter leads to the
following constraint on the parameters of the model:
\be
(M_D^\dagger M_D)_{11} \simeq m_0^2,
\label{dm1}
\ee
where $m_0 = {\cal O}(0.1)$ eV. In terms of Yukawa couplings and
mixing angles it reads: 
\be
 f_1^2 \, c_{R12}^2 \, c_{R13}^2 
 +
 f_2^2 \, c_{R13}^2 \, s_{R12}^2 
 + 
 f_3^2 \, s_{R13}^2  
 \simeq 3.3 \cdot 10^{-25}.
\label{dm2}
\ee
In other words, at least one of the couplings $f_i$  must be of order
$6\cdot 10^{-13}$. We choose it to be $f_1$. Though being
much smaller than the Yukawa constants in the charged lepton sector,
this value does not contradict to anything and is stable against loop
corrections in the $\nu$MSM.

The dark matter constraint allows us to fix the absolute values of
the active neutrino masses as follows: the mass of the lightest
active neutrino should lie in the range $m_1 \le m_{\nu}^{\rm dm}
={\cal O}(10^{-5})$~eV, while other masses are $m_3 =
[4.8^{+0.6}_{-0.5}] \cdot 10^{-2}$ eV and $m_2=
[9.05^{+0.2}_{-0.1}]\cdot 10^{-3}$eV ($[4.7^{+0.6}_{-0.5}]\cdot
10^{-2}$eV) in the normal (inverted) mass
hierarchy~\cite{Asaka:2005an}.

There are further constraints on Yukawa couplings coming from Big
Bang Nucleosynthesis (BBN) and from the absolute values of neutrino
masses. Since the second and third active neutrinos are much heavier
than the first, one must have $f_{2,3} \gg f_1$. This means that
these neutrinos equilibrate before BBN and  spoil its predictions,
unless they decay before BBN.  This leads to the constraint 
\cite{Akhmedov:1998qx} $M_{2,3} > 1$ GeV, and correspondingly, to
$f_2^2 > 3\cdot 10^{-16},~f_3^2 > 2 \cdot 10^{-15}$, where the
specific numbers were derived assuming small mixing angles 
$\theta_{Lij}$ and normal hierarchy. Together with eq. (\ref{dm2}),
this tells that the mixing angles $\theta_{R12}$ and $\theta_{R13}$
must be small, with 
\be
s_{R12}< 3.3 \cdot 10^{-5}\left(\frac{\mbox{GeV}}{M_2}\right)^{1/2}
~~ \mbox{and}~~ 
s_{R13}< 1.4 \cdot 10^{-5}\left(\frac{\mbox{GeV}}{M_3}\right)^{1/2}.
\label{eq:Const_SR}
\ee
For the case of inverted hierarchy the conclusion on the smallness of
$s_{R12}$ and $s_{R13}$ is the same.

We would now like to see whether this model can account for the
baryon asymmetry of the universe in a specified parameter range.

{\em The $\nu$MSM and baryon asymmetry.}--- 
In considering the problem of the baryon asymmetry we take the most
conservative point of view and assume the validity of the standard
Big Bang theory well above the electroweak scale. Since the
right-handed neutrino Yukawa coupling constants are small, $N$'s are
out of thermal equilibrium at high temperatures and may come into
thermal equilibrium at smaller temperatures. In any case, the
lightest sterile neutrino, playing the role of dark matter, never
equilibrates. Because of this we will assume that initial
concentrations of right-handed neutrinos are equal to zero.

To the best of our knowledge there has been only one study of baryon
production in the model (\ref{lagr}) with small Majorana
masses\footnote{Note that the Majorana neutrinos with masses much
larger than the electroweak scale give rise to thermal leptogenesis
\cite{Fukugita:1986hr}.}. Namely, in  ref. \cite{Akhmedov:1998qx}
Akhmedov, Rubakov and Smirnov (ARS in what follows) proposed an
interesting idea that the baryon asymmetry can be generated through
CP-violating sterile neutrino oscillations. For small Majorana masses
the total lepton number of the system, defined as the lepton number
of active neutrinos plus the total helicity of sterile neutrinos, is
conserved and equal to zero during the universe's evolution. However,
because of oscillations the lepton number of active neutrinos becomes
different from zero and gets transferred to the baryon number due to
rapid sphaleron transitions \cite{Kuzmin:1985mm}. Roughly speaking,
the resulting baryon asymmetry is equal to the lepton asymmetry at
the sphaleron freeze-out \cite{Kuzmin:1987wn}, with the exact
relation  in \cite{Khlebnikov:1996vj}. 

According to the ARS computation, the  produced lepton asymmetry is
entirely expressed through the Majorana masses $M_I$ of sterile
neutrinos, Yukawa couplings $f_i$ and the mixing matrix $K_R$ and can
easily be of the required order of magnitude. However, the dark
matter and BBN constraints put  severe limitations on the phase space
of the $\nu$MSM and we found that no choice of the parameters,
consistent with (\ref{dm2},\ref{eq:Const_SR}), can lead to the
observed baryon asymmetry of the universe, if the ARS equations are
used. This can be understood in the following way. The CP-violating
effects must be proportional to the Jarlskog determinant
\cite{Jarlskog:1985ht} related to the matrix $K_R$, $J = c_{R13}^2
c_{R12}c_{R23}s_{R13} s_{R12}s_{R23} \sin\delta_R$. With the
constraints on the mixing angles discussed above, the value of $J$ is
at most $10^{-10}$. Other factors, such as the total number of
degrees of freedom, make the asymmetry predicted by ARS formula well
below the observed value in the region of parameter space we are
interested in. However, we have reached a different conclusion. Below
we will reconsider the neutrino oscillations in the early universe
and identify the similarities and crucial differences with ARS.

In general terms, the evolution of possible lepton asymmetries 
can be found with the use of kinetic equations for a complete
neutrino density matrix \cite{Dolgov:1980cq} - \cite{Sigl:1992fn}. In
our case this is a $12\times 12$ matrix with components describing
the mixing of active neutrinos and anti-neutrinos as well as the
sterile neutrinos of different helicity states:
\be
\rho=
\left(
\begin{array}{c c c c}
\rho_{LL} &  \rho_{L\bar L}& \rho_{LN} &\rho_{L\bar N}\\
\rho_{\bar L L} &  \rho_{\bar L\bar L}& \rho_{\bar L N} &\rho_{\bar L\bar N}\\
\rho_{N L} &  \rho_{N\bar L}& \rho_{N N} &\rho_{N\bar N}\\    
\rho_{\bar N L} &  \rho_{\bar N\bar L}& \rho_{\bar N N} &\rho_{\bar N\bar N}\\    
\end{array}
\right)~,
\ee
where different entries in $\rho$ are the $3\times 3$ matrices
describing the neutrino states in different sectors (by $\bar N$ we
denote the negative helicity state of the sterile neutrinos). This
density matrix satisfies the kinetic equation \cite{Sigl:1992fn}
which can be written in the form
\be
i\frac{d\rho}{dt} =
[H,\rho]-\frac{i}{2}\{\Gamma,\rho\}+\frac{i}{2}\{\Gamma^p,1-\rho\}~,
\label{cineq}
\ee
where $H=k(t)+H^0+H_{int}$ is the Hermitian effective Hamiltonian
incorporating the medium effects on neutrino propagation \cite{MSW},
$k(t)\sim T$ is the neutrino momentum, $\Gamma$ and $\Gamma^p$ are
the destruction and production rates correspondingly. Following ARS,
we will use the Boltzmann statistics and replace the last term in
(\ref{cineq}) by $i \Gamma^p$. 

If the system is in thermal equilibrium, the equilibrium density
matrix, given by $\rho^{eq} = \exp{(-H/T)}$ must satisfy eq.
(\ref{cineq}), which gives the relation  $ \Gamma^p =
\frac{1}{2}\{\Gamma,\rho^{eq}\}$ \footnote{For numerical estimates
$\rho^{eq}$ can be safely replaced by $\exp{(-k/T)}$, as was done in
ARS analysis, since the corrections to $\rho^{eq}$ coming from $H^0$ and
$H_{int}$ are small.}.

In the leading approximation (all Yukawa couplings $f_i$ are
neglected) the Hamiltonian $H^0$ is diagonal and can be written as
\be
H^0=\mbox{diag}(H_{LL}^0,H_{\bar L\bar L}^0,H_{N N}^0,H_{\bar N\bar N}^0)
\ee
with, in turn \cite{Weldon:1982bn},
\be
H_{LL}^0 = H_{\bar L\bar L}^0 =
 \frac{T^2}{k(t)}
 \left[ \frac{3 g_W^2 + g_Y^2 }{32}   \mbox{diag}(1,1,1) +
\frac{1}{8}\mbox{diag}(h_e^2,h_\mu^2,h_\tau^2) \right]~,
\ee
where the first term comes from the electroweak gauge correction to
the active neutrino propagator, and the second from the Higgs
correction. $h_e^2,~h_\mu^2$ and $h_\tau^2$ are the charged lepton
Yukawa couplings. For the sterile neutrinos we have
\be
H_{N N}^0=H_{\bar N\bar N}^0
=\frac{1}{2k(t)}\mbox{diag}(M_1^2,M_2^2,M_3^2)~.
\ee

Because of the structure of $H^0$ the consideration of oscillating
neutrinos can be simplified significantly. To remove the trivial time
dependence of the density matrix, make a change
\be
\rho= U(t) \tilde \rho U^\dagger(t)~,
\ee
where $U= \exp{(-i\int_0^t dt' H^0(t'))}$, which brings us to the
evolution equation in the interaction picture. Now, as we will see
later (and in accordance with ARS) the most important region for
leptogenesis is $T_L \simeq (\Delta M^2 M_0)^{\frac{1}{3}}$, where
$\Delta M^2$ is the typical value of the Majorana mass squares and
$M_0 \simeq 7 \cdot 10^{17}$ GeV  appears in the time-temperature
relation, $t = \frac{M_0}{2 T^2}$. At this time most of the off-diagonal
elements of density matrix $\tilde\rho$ undergo very rapid
oscillations because of $U(t)$ and decouple from the system. These
elements are listed below: $\rho_{LN},~ \rho_{L\bar N},~ \rho_{\bar L
N},~ \rho_{\bar L\bar N},~\rho_{N L},~\rho_{N\bar L},~ \rho_{\bar N
L},~\mbox{and}~ \rho_{\bar N\bar L}$; they can be safely put to zero.
In addition, all non-diagonal parts of $\rho_{LL}$ and $\rho_{\bar
L\bar L}$ can be neglected as well. In physics terms, the
oscillations between $L$ and $N$ are strongly suppressed since these
particles have very different effective masses. The same is true for
$L_\alpha \rightarrow L_\beta, ~\alpha \neq \beta$ oscillations. On
the other hand, the non-diagonal parts of $\rho_{N N}$ and  $\rho_{\bar
N\bar N}$ must be kept, as the corresponding exponentials in $U(t)$ are 
of order 1. Another four non-diagonal  entries of
the density matrix, $\rho_{L\bar L},~\rho_{\bar L L},~ \rho_{N\bar
N},~\mbox{and}~\rho_{\bar N N}$ can be removed as they include the
processes with lepton number nonconservation in the active sector
and  helicity-flip in the sterile sector, suppressed by the small
Yukawa couplings and small mass to temperature ratio for $N$'s
\cite{Akhmedov:1998qx}. As a result, the kinetic equation contains
the density matrices $\rho_{N N}$ and $\rho_{\bar N\bar N}$ where
coherent quantum effects are essential, and the diagonal parts of
$\rho_{LL}$ and $\rho_{\bar L\bar L}$, describing concentrations of
active neutrinos and anti-neutrinos. The latter quantities were not
considered by ARS.

\begin{figure}
\epsfysize=5cm
\centerline{\epsffile{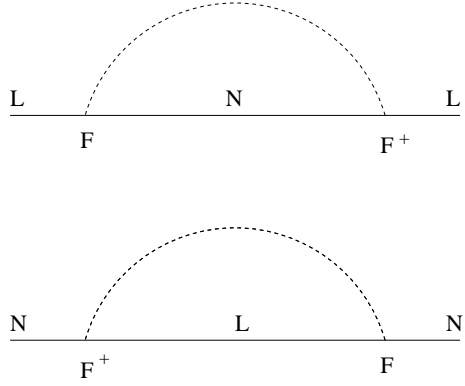}}
\caption{The diagrams contributing to the real potential and rate
$\Gamma$ in the kinetic equation.}
\label{diagr}
\end{figure}

Now we are ready to write our kinetic equations. We will do that for
$L$ and $N$ only, as the equations for $\bar L$ and $\bar N$ are given
by CP-conjugation. The medium effects follow from computation of
real and imaginary parts of the two-point Greens functions on Fig.
\ref{diagr}\footnote{Note that the absorptive parts of these diagrams
are in fact not suppressed by the Majorana neutrino masses, as was
stated by ARS, but by the effective mass of the Higgs boson. However,
this effect leads to the same numerical estimates of the absorption
rates as in \cite{Akhmedov:1998qx} (see \cite{AS}).}.
%%%%%%%%%%%%%%%%%%%%%%%%%%%%%%%%%%%
%%%%%%%%%%%%%%%%%%%%%%%%%%%%%%%%%%%%%
The equation for the sterile neutrinos is
\be
i \frac{d\tilde \rho_{N N}}{dt}= [H_{int}^N(t),\tilde \rho_{N N}]
-\frac{i}{2}\{\Gamma^N(t),\tilde \rho_{N N}-\tilde \rho_{N N}^{eq}\}
+\frac{\sin\phi}{8} \, T \,
U^\dagger(t) F^\dagger(\rho_{LL}-\rho_{LL}^{eq})F U(t)~.
\label{NN}
\ee
For the diagonal part of the active neutrino density matrix we have:
\be
i \frac{d \rho_{LL}}{dt}=
\mbox{diag} \left[ \frac{\sin\phi}{8} \, T \,
F U(t)(\tilde \rho_{N N}-\tilde \rho_{NN}^{eq})U^\dagger(t)F^\dagger~
-\frac{i}{2}\{\Gamma^L(t), \rho_{LL}- \rho_{LL}^{eq}\}\right]~.
\label{LL}
\ee
Here the following notation is introduced:
\be
H_{int}^N(t) = \frac{T}{8} U^\dagger(t) K_R f_d^2 K_R^\dagger U(t),~~
\Gamma^N(t) = \sin\phi H_{int}^N(t),~~
\Gamma^L(t)=\frac{\sin \phi}{8} \, T \, K_L f_d^2 K_L^\dagger ~,
\ee
where $\tilde\rho_{NN}^{eq}$ and $\rho_{LL}^{eq}$ are the diagonal
equilibrium density matrices for sterile and active neutrinos
respectively, and $\sin\phi \simeq 0.02$ \cite{Akhmedov:1998qx} is
the ratio of the absorptive part of the diagrams to their real part.
These equations are the basis for the analysis of the time evolution
of the asymmetries as a function of the parameters of the $\nu$MSM.

The equation (\ref{NN}) with the third term omitted coincides with
that of ARS. The first term in (\ref{NN}) describes the oscillations
of the sterile neutrinos, whereas the second is responsible for
absorption and creation of them. The first term in (\ref{LL})
describes the transfer of leptonic number from sterile neutrinos to
the active ones. It comes from the first diagram of Fig. \ref{diagr}
and incorporates the change in $\Gamma$ due to the presence of the
CP-breaking medium, coming from non-trivial $\tilde \rho_{N N}$. The
trace of the second term of (\ref{NN}) is exactly equal to the trace
(with a minus sign) of the first term in (\ref{LL}), ensuring the
exact conservation of the lepton number. The second term in
(\ref{LL}) describes the absorptive processes with active neutrinos
and ensures that the system eventually thermalizes. The third term in
(\ref{NN}) is a counterpart of this one and takes into account the
transfer of the active lepton number into the sterile sector. It
comes from the change in $\Gamma$ due to the presence of asymmetries
in active neutrinos.

Our equations have a rich CP-violating structure. In addition to the
CP-violating phase in the $K_R$ matrix, they contain 3 extra phases,
$\delta_L$ and $\alpha_1,~\alpha_2$. Note that the Majorana phases
$\beta_1,~\beta_2$ do not appear in the limit of the small Majorana
masses we are interested in. The equations (\ref{NN},\ref{LL}) are to
be solved with initial condition  $\rho_{LL}-\rho_{LL}^{eq}=0|_{t=0}$
(no lepton-flavour asymmetry at the beginning) and $\tilde \rho_{N
N}=0|_{t=0}$ (no sterile neutrinos in the initial state).

For sufficiently small Yukawa couplings, when all sterile neutrinos
are still out of thermal equilibrium at the freeze-out of the
sphaleron transitions, the solution can be found perturbatively. This
is realized provided $f_i^2 < 2\times 10^{-14}$ 
\cite{Akhmedov:1998qx}. This regime requires the relatively light
sterile neutrinos, as from the see-saw formula, and with the use of
the absolute values of the active neutrino masses one gets $M_{2,3} <
12$ GeV. For these masses the processes with lepton number
non-conservation induced by the Majorana neutrino masses are safely
out of thermal equilibrium.

For the higher Majorana masses and larger Yukawa couplings the system
can be solved numerically. It is clear, however, that if both $N_C$
and $N_B$ are already thermalized at the sphaleron freeze-out, no
baryon asymmetry is produced, since the production of the $N_D$ is
highly suppressed and at that time no neutrino can carry the lepton
asymmetry. This leads to the requirement that all the sterile
neutrino masses must be smaller than the electroweak scale, exactly
in accordance with our definition of the $\nu$MSM.

Since the aim of our paper is not the analysis of the complete
parameter space of the $\nu$MSM but rather the existence proof of a
possibility of simultaneous explanation of neutrino oscillations,
dark matter and baryon asymmetry, we will consider here the
perturbative regime only (we checked, however, that numerical
solution coincides with the perturbative one in the limit of its
validity).

In the region of parameter space where the dark matter constraint is
satisfied the generation of the lepton asymmetry occurs as follows.
First, the sterile neutrinos are created  in a CP-invariant state
(the Jarlskog determinant is very small for $K_R$!) due to the second
term in eq. (\ref{NN}). However, their density matrix is non-trivial
and contains non-diagonal terms. Because of that and because of the
presence of CP-violating phases in $\tilde K_L$, this leads to
generation of CP-asymmetries in the active neutrino flavours, due to
the first term in (\ref{LL}).  Finally, this asymmetry is partially
transferred into the total active (or sterile, with the opposite
sign) asymmetry. 

Now we are ready to solve eqns. (\ref{NN}) and (\ref{LL}) with the
use of perturbation theory in $f_i^2$. For this end they can be
rewritten in the integral form, as 
\begin{eqnarray}
\tilde\rho_{NN} &=& - i \int_0^t dt' \,
(\mbox{right-hand-side of eq.(\ref{NN}))}~,\\ 
\rho_{LL} &=&
\rho_{LL}^{eq} -i \int_0^t dt' \, (\mbox{right-hand-side of eq.(\ref{LL}))}
\end{eqnarray}
and then solved iteratively. We will present below the leading
contributions only and refer for the details of calculation to \cite{AS}.

Let us first discuss the asymmetries in active neutrino flavours
induced by the non-thermal density matrix of sterile neutrinos. This
effect comes from the first term in (\ref{LL}) and results in a
second order effect in $\Delta_{L} = \rho_{LL}-\rho_{\bar L \bar L}$:
\begin{eqnarray}
  \frac{d \Delta_{L}}{dt} = \frac{\sin \phi}{8} \, T  \,
 \mbox{diag} 
 \left(F \rho_{NN} F^\dagger - F^\ast \rho_{\bar N \bar N} F^T \right)\,.
\end{eqnarray}
To leading order in perturbation theory the elements of
$\rho_{NN}$ come from eq. (\ref{NN}) by neglecting the 
third term on the right-hand side. 

As in the ARS analysis, the asymmetries are produced most effectively
at the typical temperature  $T_L \simeq (\Delta M^2 M_0)^{1/3}$ and
take constant values in the later evolution. Then, when $T_L \gg T_W
\sim 100$ GeV, we obtain the following expression
\begin{eqnarray}
  (\Delta_L)_{\alpha \alpha} |_{T_W}
  = 
  \frac{\pi^{\frac 32} \sin^2 \phi }
  { 12 \cdot 3^{\frac{1}{3}} \Gamma (\frac{5}{6})  } \,
  \, \sum_{I>J} \delta^{\alpha}_{IJ} \, 
  \frac{M_0{}^{\frac{4}{3}}}{(\Delta M_{IJ}^2)^{\frac{2}{3}}} \,,
  \label{DeltaL}
\end{eqnarray}
where the effective CP violation parameter is given by
\begin{eqnarray}
  \delta^{\alpha}_{IJ} = \mbox{Im} \left[ F_{\alpha I} \, 
    (F^\dagger F)_{IJ} F^\dagger_{J \alpha} \right] \,.
\end{eqnarray}
It can be seen that the total lepton number of active neutrinos is
zero, i.e. $\delta^{e}_{IJ} + \delta^{\mu}_{IJ} + \delta^{\tau}_{IJ}=0$. 
Since the mixing angles $s_{R12}$ and
$s_{R13}$ are very small, as shown  in (\ref{eq:Const_SR}), we can
safely put them to zero in what follows, $s_{R12} = s_{R13} = 0$.  In
this case we obtain
\begin{eqnarray}
  (\Delta_L)_{\alpha \alpha} |_{T_W}
  = a^\alpha \, 
  \frac{\pi^{\frac 32} \, \sin^2 \phi }
  { 48 \cdot 3^{\frac{1}{3}} \Gamma (\frac{5}{6})  } \,
  \frac{f_2 f_3 ( f_3^2 - f_2^2 )M_0{}^{\frac{4}{3}}}
  {(\Delta M_{32}^2)^{\frac{2}{3}}} \,,
\end{eqnarray}
where $a^\alpha$ are given by
\begin{eqnarray}
  a^{e}
  &=&
  - 4 s_{R23} c_{R23} 
  \left[
     s_{L12} s_{L13} c_{L13} \, \sin (\delta_L + \alpha_2)
  \right] \,,
  \nonumber \\
  a^\mu
  &=&
  4 s_{R23} c_{R23} 
  \left[
     s_{L12} s_{L13} c_{L13} s_{L23}^2 \, \sin (\delta_L + \alpha_2)
    - 
     c_{L12} c_{L13} s_{L23} c_{L23} \, \sin \alpha_2
  \right] \,,
  \nonumber \\
  a^\tau
  &=&
  4 s_{R23} c_{R23} 
  \left[
     s_{L12} s_{L13} c_{L13} c_{L23}^2 \, \sin (\delta_L + \alpha_2)
    + 
     c_{L12} c_{L13} s_{L23} c_{L23} \, \sin \alpha_2
  \right] \,.
\end{eqnarray}
It is clearly seen that the phases $\delta_L$ and $\alpha_2$ in the
mixing matrix $\widetilde K_L$ are crucial for the asymmetry in the
active neutrino flavours\footnote{The CP-violating phases which can
be found experimentally from active neutrino oscillations depend also
on $\beta_2$ and thus cannot be used to fix uniquely  $\delta_L$ and
$\alpha_2$.}. Furthermore, it should be noted that the observed large
mixing angles in the atmospheric and solar neutrino oscillations
indicate that these parameters are not suppressed but can be ${\cal
O}(1)$. As an illustration of numerics, we shall take the typical
values of the Dirac Yukawa couplings to be
\begin{eqnarray}
  \label{eq:Typical_Yukawa}
  f_2{}^2 = \frac{  m_{\rm sol} M_2 }{ v^2 } \,,~~~
    f_3{}^2 = \frac{  m_{\rm atm} M_3 }{ v^2 } \,,~~~
\end{eqnarray}
where $m_{\rm sol} = \sqrt{ \Delta m_{\rm sol}^2} \simeq 9.1$ meV and
$m_{\rm atm} = \sqrt{ \Delta m_{\rm atm}^2} \simeq$ 51 meV. Then the
asymmetries are:
\begin{eqnarray}
  (\Delta_L)_{\alpha \alpha}|_{T_W}
  &=& 
  a^{\alpha} \,
  \frac{\pi^{\frac 32} \, \sin^2 \phi  }
  { 48 \cdot 3^{\frac{1}{3}} \Gamma (\frac{5}{6})  } \,
  \frac{m_{\rm sol}^{\frac 12} m_{\rm atm}^{\frac 32} M_2^{\frac{1}{2}} 
     M_3^{\frac{3}{2}}  M_0{}^{\frac{4}{3}}}
  { v^4 \, (\Delta M_{32}^2)^{\frac{2}{3}}} \,,
  \nonumber \\
  &\simeq&  10^{-6} \, a^\alpha 
  \left( \frac{ 10^{-6} }{\Delta M_{32}^2/M^2_3} \right)^{\frac{2}{3}}
  \left( \frac{ M_3 }{10 \GeV } \right)^{\frac{2}{3}} 
  \label{eq:Delta_L}
  \,.
\end{eqnarray}
This equation shows that sizable flavour asymmetries can be 
generated when the heavier ``bright'' and ``clear''  neutrinos  are
highly degenerate in mass. In fact, these flavour asymmetries are
partially converted into the baryon asymmetry of the universe even if
the total lepton asymmetry is zero, due to effects associated
with the masses of the charged leptons \cite{Kuzmin:1987wn} (for an
exact computation of the mass corrections see \cite{Laine:1999wv}).
However, the conversion rate is suppressed by the factor
$\simeq\frac{5}{13\pi^2}\frac{m_\tau^2}{T^2}$ and asymmetries of
$(\Delta_L)_{\tau \tau} \sim {\cal O}(0.01)$ are required to
account for the observed baryon asymmetry, if this mechanism is
used.

However, this is not the end of the story.  The above asymmetries in
the active neutrino flavours trigger the generation of asymmetries in
the sterile neutrino sector in a very efficient way. This phenomenon
is due to the third term in (\ref{NN}). Indeed, the asymmetries
$\Delta_N = \rho_{NN}-\rho_{\bar N \bar N}$  are generated as
\begin{eqnarray}
  \frac{d \Delta_N}{dt} = \frac{\sin \phi}{8} \, T \,
  ( F^\dagger \, \Delta_L \, F ) \,.
\end{eqnarray}
Since $\Delta_L$ takes constant values after the production time, this
equation can be easily solved.  At the electroweak temperature the
asymmetries in the sterile neutrinos are 
\be
  (\Delta_N)_{II}|_{T_W} = \frac{\sin \phi}{8} \frac{M_0}{T_W} 
  ( F^\dagger \, \Delta_L |_{T_W} \, F)_{II} ~,
  \ee 
where $\Delta_L |_{T_W}$ is defined in (\ref{DeltaL}). We should
stress that these asymmetries are generated at higher order in
$f_i^2$ compared with $\Delta_L$ (third versus second order); 
however the production processes continue below the temperature $T_L$
and thus receive a huge enhancement factor~\footnote{
  Note that when $M_0/T_W \rightarrow \infty$ the perturbative
  approximation breaks down and one must solve the equations exactly.
  It is clear, however, what happens in this limit: the system will
  eventually equilibrate because of the second terms in
  (\ref{NN},\ref{LL}) and all asymmetries will go away. However,
  before this time the effect is accumulating, as all sterile
  neutrinos are out of thermal equilibrium.}
of the order of $M_0/T_W$.
Since the ``dark'' sterile neutrino
only possesses small Yukawa couplings as shown in (\ref{dm1}), the
asymmetries are generated in $(\Delta_N)_{22}$ and
$(\Delta_N)_{33}$. 

Let us estimate the trace of $\Delta_N$, which is crucial for the
baryon asymmetry of the universe, in the specific parameter choice 
discussed above.  We find from (\ref{eq:Delta_L}) that
\begin{eqnarray}
  \mbox{Tr} \Delta_N|_{T_W}
  = \frac{\pi^{\frac 32} \, \sin^3 \phi}
  { 384 \cdot 3^{\frac{1}{3}} \Gamma (\frac{5}{6})  } \,
  \sum_{\alpha,I} a^{\alpha} \abs{F_{\alpha_I}}^2 \,
  \frac{m_{\rm sol}^{\frac 12} m_{\rm atm}^{\frac 32} M_2^{\frac{1}{2}}
     M_3^{\frac{3}{2}}  M_0{}^{\frac{7}{3}}}
  {v^4 \, T_W \, (\Delta M_{32}^2)^{\frac{2}{3}}} \,.
\end{eqnarray}

Further simplifications come about as we notice that the successful
asymmetry generation requires a significant mass degeneracy in $M_2$
and $M_3$ (see below).  In order to explain the observed active neutrino 
mass pattern in the normal hierarchy, considered below for numerical
estimates, one should have a certain  hierarchy in Yukawa couplings
$f_3{}^2 \gg f_2{}^2$ ($\gg f_1{}^2$), which gives at the leading
order
\begin{eqnarray}
  \sum_{\alpha,I} a^{\alpha} \abs{F_{\alpha_I}}^2 = 
  f_3{}^2 \, \delta_{\rm CP} \,,
\end{eqnarray}
where 
\begin{eqnarray}
  \delta_{\rm CP}
  &=&
  4 s_{R23} c_{R23} 
  \Bigl[
  s_{L12} s_{L13} c_{L13} 
  \bigl( ( c_{L23}^4 + s_{L23}^4 ) c_{L13}^2 - s_{L13}^2 \bigr)
  \cdot \sin (\delta_L + \alpha_2)
  \nonumber \\
  &&~~~~~~~~~~~~~~~+
    c_{L12} c_{L13}^3 s_{L23} c_{L23} \, ( c_{L23}^2 - s_{L23}^2 )
   \cdot
   \sin \alpha_2 
  \Bigr] \,.
\end{eqnarray}
We can see that the CP violation parameter $\delta_{CP}$ can be
${\cal O}(1)$.  Finally, the total asymmetry in the sterile neutrino
sector  is given by
\begin{eqnarray}
  \mbox{Tr} \Delta_N|_{T_W}
  = \delta_{\rm CP} \, \frac{\pi^{\frac 32} \, \sin^3 \phi }
  { 384 \cdot 3^{\frac{1}{3}} \Gamma (\frac{5}{6})  } \,
  \frac{m_{\rm sol}^{\frac 12} m_{\rm atm}^{\frac 52} M_2^{\frac{1}{2}}
     M_3^{\frac{5}{2}}  M_0{}^{\frac{7}{3}}}
  {v^6 \, T_W \, (\Delta M_{32}^2)^{\frac{2}{3}}} \,.
\end{eqnarray}

The total lepton number conservation tells us that
$\mbox{Tr}\Delta_N|_{T_W} + \mbox{Tr}\Delta_L|_{T_W} = 0$ and,
therefore, that the lepton asymmetry in the active neutrino sector is
generated \footnote{Equally, the generation of a net lepton number in
the sector of active neutrinos can be understood as follows. The
asymmetries (\ref{DeltaL}) in active neutrino flavours are the
subject of dissipation described by the second term in (\ref{LL}). Since
$\Gamma^L$ does depend on neutrino flavour, these asymmetries
evolve differently, leading to the total lepton asymmetry.}. 
It is partially converted into the baryon asymmetry due
to the rapid sphaleron conversion as $\Delta B = - \frac{28}{79} \,
\mbox{Tr}\Delta_L|_{T_W} = + \frac{28}{79} \,
\mbox{Tr}\Delta_N|_{T_W}$ \cite{Kuzmin:1987wn,Khlebnikov:1996vj}. 
This completes the computation of the baryon asymmetry of the
universe in the $\nu$MSM satisfying the dark matter constraint. 
Accounting for the entropy factor at the electroweak temperature, the
baryon to entropy ratio is obtained as
\begin{eqnarray}
  \frac{n_B}{s} = 7 \cdot 10^{-4} \,\mbox{Tr} \Delta_N|_{T_W}
  = 2 \cdot 10^{-10} \, \delta_{\rm CP} \,
  \left( \frac{ 10^{-6} }{\Delta M_{32}^2/M^2_3} \right)^{\frac{2}{3}}
  \left( \frac{ M_3 }{10 \GeV } \right)^{\frac{5}{3}} \,.
\end{eqnarray}
This shows that the correct baryon asymmetry of the universe
$\frac{n_B}{s} \simeq (8.8-9.8)\times 10^{-11}$ is generated when the
heavier sterile neutrinos with the masses, say, $10$ GeV are
degenerate to one part in $10^6$.  This looks like a strong fine
tuning but may also indicate the intriguing mass relation $|M_3-M_2|
\sim M_1$.  In this mass range the decays of heavier sterile neutrinos
induce no significant entropy dilution.
The leptogenesis temperature for degenerate neutrinos is rather low,
$T_L \sim 10^{4}$ GeV.  Note that with this low temperature the right-handed
electron is always thermalized~\cite{Campbell:1992jd} so that
the equilibrium formulas of~\cite{Kuzmin:1987wn,Khlebnikov:1996vj} can be used.
Moreover,  if the
inverted hierarchy pattern is chosen for the active neutrino masses,
even stronger degeneracy is needed, since an extra suppression
of the baryon asymmetry is coming from the small difference between
$f_2$ and $f_3$, required in this case.

{\em Conclusions.}--- 
Let us summarize the obtained results. The $\nu$MSM with three
right-handed neutrinos with masses smaller than the electroweak scale
is the simplest and the most economical extension of the Minimal
Standard Model. It shares with the MSM its advantages
(renormalisability and agreement with most particle physics
experiments) and its fine-tuning problems (the gauge hierarchy
problem, flavour problem, etc). However, unlike the MSM, the
$\nu$MSM  can explain simultaneously three different  phenomena,
observed experimentally, namely neutrino oscillations, dark matter,
and baryon asymmetry of the universe. The parameter-space of the
model is rather constrained (the heavier neutrinos are required to be
quite degenerate in mass, the Yukawa coupling $f_1$ and mixing angles
$\theta_{R12}$ and $\theta_{R13}$ must be very small), which,
however, makes this model experimentally testable. The analysis of
possible experimental signatures of the $\nu$MSM goes beyond the
scope of the present letter\footnote{It would be interesting to see,
for example, whether the $\nu$MSM can also explain in the suggested
parameter range the LSND anomaly \cite{deGouvea:2005er} and the
pulsar kick velocities \cite{Kusenko:1997sp}.}. We would just like
to mention that on the astrophysical side, the best signal would be
the $\gamma$ radiation coming from the decay of dark neutrinos. On
the particle physics side, as all the sterile neutrinos are
relatively light, one could imagine that all parameters of the
$\nu$MSM, in particular the CP-violating phases, will be determined
one day (this is clearly a very hard, if not impossible experimental
challenge because of the smallness of the Yukawa couplings). One
would then be able to test the baryogenesis formula and in particular
its sign.  

%\newpage

%%%%%%%%%%%%%%%%%%%%%%%%%%%%%%%%%%%%%%%%%%%%%%%%%%%%%%%%%%%%%%%%%%%%%%%%
% Acknowledgments
%%%%%%%%%%%%%%%%%%%%%%%%%%%%%%%%%%%%%%%%%%%%%%%%%%%%%%%%%%%%%%%%%%%%%%%%
{\em Acknowledgments.}---
This work has been supported by the Swiss Science Foundation and by
the Tomalla foundation. We thank Steve Blanchet for helpful
discussions and Stephen Davis for careful reading of the manuscript.
We are grateful to Evgenii Akhmedov, Valery Rubakov and Alexey Smirnov
for correspondence and valuable comments.

%%%%%%%%%%%%%%%%%%%%%%%%%%%%%%%%%%%%%%%%%%%%%%%%%%%%%%%%%%%%%%%%%%%%%%%%
% References
%%%%%%%%%%%%%%%%%%%%%%%%%%%%%%%%%%%%%%%%%%%%%%%%%%%%%%%%%%%%%%%%%%%%%%%%
%\input{refs.tex}

\end{document}